\documentclass[structabstract]{aa}  
\usepackage{graphicx}
\usepackage{txfonts}
\usepackage[authoryear]{natbib}
\begin{document}
   \title{The dynamic atmospheres of Mira stars: comparing the CODEX models to PTI time series of TU\,And}

   \author{M. Hillen\inst{1}
          \and
          T. Verhoelst\inst{1,2}
	  \and
          P. Degroote\inst{1}\thanks{Postdoctoral Fellow, Fund for Scientific Research of Flanders (FWO)}
          \and
          B. Acke\inst{1 \star}
          \and
          H. van Winckel\inst{1} 
          }

   \institute{Instituut voor Sterrenkunde (IvS), K.U.Leuven,
              Celestijnenlaan 200D, B-3001 Leuven, Belgium\\
              \email{Michel.Hillen@ster.kuleuven.be}
         \and
             Belgian Institute for Space Aeronomy, Brussels, Belgium      
             }

   \date{Received ? ?, 2011; accepted ? ?, 2011}
   \authorrunning{Hillen et al.}
   \titlerunning{Comparing the CODEX models to PTI time series of TU\,And}

 
  \abstract
   {Our comprehension of stellar evolution on the AGB still faces many difficulties. To improve on this, a quantified understanding of large-amplitude pulsator atmospheres 
    and interpretation in terms of their fundamental stellar parameters are essential.}
   {We wish to evaluate the effectiveness of the recently released CODEX dynamical model atmospheres in representing M-type Mira variables through a confrontation 
   with the time-resolved spectro-photometric and interferometric PTI data set of TU\,And.}
   {We calibrated the interferometric K-band time series to high precision. This results in 50 nights of observations, covering 8 subsequent pulsation cycles. 
    At each phase, the flux at 2.2~$\mu$m is obtained, along with the spectral shape and visibility points in 5 channels across the K-band. We compared the data set to
    the relevant dynamical, self-excited CODEX models.}
   {Both spectrum and visibilities are consistently reproduced at visual minimum phases. Near maximum, our observations show 
    that the current models predict a photosphere that is too compact and hot, and we find that the extended atmosphere lacks H$_2$O opacity. Since coverage in model
    parameter space is currently poor, more models are needed to make firm conclusions on the cause of the discrepancies. We argue that for TU\,And, the discrepancy 
    could be lifted by adopting a lower value of the mixing length parameter combined with an increase in the stellar mass and/or a decrease in metallicity, 
    but this requires the release of an extended model grid.}
   {}


   \maketitle
%

\section{Introduction}
In spite of their great astrophysical importance \citep[][]{Muzzin2009ApJ}, many aspects of asymptotic giant branch (AGB) stars 
remain poorly understood \citep[e.g.][and references therein]{Kerschbaum2011ASPC}. Among the AGB stars, the luminous Mira variables might be the most enigmatic 
with their long-period, large-amplitude pulsations and high mass-loss rates. In particular our understanding of their atmospheric structure, 
which is the link between enriched stellar interior and evolution-dominating wind, is at a turning point, thanks to advances on both the observational 
side (IR interferometry), and the modelling side with a new generation of self-excited atmosphere models.

The public release of four CODEX model series for M-type Mira variables by \citet[][ hereafter ISW11]{Ireland2011MNRAS}, 
greatly increases the opportunities for the general community to assess the performance of such dynamic models. Here, we present the first confrontation
of the CODEX models (see Sect.~\ref{CODEXsection}) to a fully phase-resolved 2~$\mu$m spectro-photometric and interferometric dataset covering eight pulsation cycles, 
as obtained with the Palomar Testbed Interferometer on the Mira TU\,And. This star has a spectral type M5IIIe, a progenitor mass
$M=1.15-1.4 \, M_{\odot}$, metallicity $Z=0.004-0.02$ \citep[][]{Mennessier2001A&A}, and a GCVS period P$\sim$317~d \citep[][]{Samus2009yCat}.

The most intriguing result of this comparison is that there seems to be a systematic discrepancy between the models and our observations around visual maximum: 
the star appears too hot, and its outer atmosphere too devoid of water. An identification of the source of this discrepancy is hampered by the limited coverage 
in parameter space of the models, but we discuss possible avenues of further research, and stress the need for larger published model grids and for 
observational campaigns covering the visual maximum phase in detail.


\section{Observations}
TU And was extensively observed as part of the large PTI Mira programme \citep{Thompson2002PhDT} during 60 nights from 1999 to 2005, over eight pulsation cycles.
The publicly available PTI data archive provides both squared visibility amplitudes (hereafter visibilities) and integrated photon counts. We describe the data reduction and 
calibration of these observables below.

%
\onltab{1}{
\begin{table}
\caption{The derived calibrator parameters.}             
\label{table:1}      
\centering                          
\begin{tabular}{c c c c}        
\hline\hline                 
Calibrator & SpT & $T_\mathrm{eff}$ (K) & UD Diameter (mas) \\    
\hline                        
   HD166 & K0V & 5540 $\pm$ 90 & 0.61 $\pm$ 0.05 \\      
   HD1404 & A2V & 8840 $\pm$ 175 & 0.47 $\pm$ 0.02 \\
   HD2628 & A7III & 7200 $\pm$ 50 & 0.48 $\pm$ 0.05 \\
   HD3651 & K0V & 5250 $\pm$ 50 & 0.74 $\pm$ 0.06 \\
   HD6920 & F8V & 6350 $\pm$ 200 & 0.61 $\pm$ 0.04 \\
   HD7034 & F0V & 7600 $\pm$ 450 & 0.49 $\pm$ 0.04 \\
   HD7229 & G9III & 5550 $\pm$ 150 & 0.80 $\pm$ 0.12 \\
   HD9712 & K1III & 4750 $\pm$ 240 & 0.78 $\pm$ 0.20 \\
\hline                                   
\end{tabular}
\end{table}
}

\subsection{Interferometry}
The now dismounted Palomar Testbed Interferometer (PTI) was a Y-shaped, long-baseline near-IR interferometer located at Palomar Observatory 
and operated by JPL/Caltech. It consisted of three 40 cm apertures with separations of 85-110\,m and used in pairwise 
combination \citep[see][for a detailed description]{Colavita1999ApJ}. PTI thus measured only one interferometric observable per spectral channel: the square of the ``fringe contrast'' 
or visibility amplitude. Visibilities were recorded in five channels within the K-band (2.0-2.4~$\mu $m). 
Processing of the raw data (correcting for detector biases and read-out noise, and averaging the 125s scans into fringe amplitudes) was automatically done on site, following 
procedures described in \citet{Colavita1999PASP}. Calibrating for the system visibility, i.e. the response 
of the measurement system (interferometer + atmosphere) to a point source, was done using the \textit{nbCalib} tool \citep[][]{Boden1998SPIE} provided by the NExScI which 
uses a weighted-mean formalism to estimate the system visibility at the time and position of the science observation from all calibrator 
measurements that fit the user-defined temporal and spatial constraints \citep[see][for more details]{vanBelle1999AJ}. To ensure calibration stability and consistency, 
essential for making full use of this data set as an interferometric time series, several calibrators were observed each night and were intercompared by nbCalib.

The stellar parameters
of our calibrators were determined by fitting a reddened Kurucz model atmosphere \citep[][]{Kurucz1993yCat} to high-quality archival photometry. 
The errors were estimated using Monte Carlo simulations. Table~\ref{table:1} (only available electronically) lists the calibrators and their adopted parameters. 
All calibrators have angular diameters well below the PTI resolution limit of 1~mas \citep[][]{vanBelle2005PASP} and are, on average, $10^\circ$ away from the science target. 

The calibrated visibilities of TU\,And from 50 nights were retained, having a median relative 
precision ranging from 5\% at 2.2~$\mu $m, 7.5\% at 2.4~$\mu $m and 10\% at 2.0~$\mu $m.

\subsection{Spectro-photometry} 
Together with the uncalibrated visibilities, PTI delivered the raw photon counts of each observation, typically in blocks of five observations taken
within minutes and interleaving science targets with calibrators. We derived the spectral shape over the K-band and flux-calibrated the
2.2~$\mu $m channel with a bootstrap method.

The spectral shape calibration procedure consists of dividing the science-target normalised photon counts by the ones of the calibrator and multiplying the result
with the intrinsic spectral shape of the calibrator. To obtain an absolute flux in the 2.2~$\mu $m channel, one calibrator measurement on either side of the science observation
with respect to airmass is required to solve for the atmospheric extinction per unit airmass $A_v = (\log[I_2/I_1]+\log[S_1/S_2])/(m_2-m_1)$ and the 
instrument spectral response $R_v = S_1/(I_1 \times \exp(-A_v m_1))$, with $I_i$ the intrinsic and $S_i$ the measured flux, and 
$m_i$ the airmass, of the $i$th calibrator. The calibrated flux is then calculated as $I_{sci} = S_{sci}/R_v \times \exp(A_v m_{sci})$, where $S_{sci}$ denotes the
measured photon counts and $m_{sci}$ the airmass of the science target. For each observing block, the above procedure is applied 5000 times to a randomly picked 
science observation, corresponding random calibrator observations and calibrator intrinsic fluxes. Finally, the distributions of all observing blocks within a single night are 
co-added, and the average and standard deviation are taken from the resulting normal distribution, as the final calibrated flux and its error.

To obtain the calibrator intrinsic fluxes, the PTI optical configuration was taken into account: after beam combination and dispersion by a prism, the wavefront was sent
through a broadband K Barr filter onto the detector pixels \citep[][]{Colavita1999ApJ}. Each of the Kurucz models in the Monte Carlo distributions obtained with the spectral energy
distribution fitting, is converted into the effective flux I in each of the PTI channels as:
\begin{equation}\label{equation1}
 I = \frac{\int_{\lambda_l}^{\lambda_u} [(M \cdot F) \otimes G] \lambda d \lambda}{\int_{\lambda_l}^{\lambda_u} [F \otimes G] \lambda d \lambda}
\end{equation}   
with $M$ the Kurucz model, $F$ the filter profile, $G$ a Gaussian of $FWHM = 0.7 \, \mu$m (M. Colavita, private communication), and $\lambda_l$, $\lambda_u$ the edges of the 
spectral channel under consideration. 

Finally, we constructed a visual light curve by retrieving observed V-band photometry from the AAVSO and AFOEV. Visual phases (hereafter simply phases) were computed by
fitting a sine to ten cycles around the times of the PTI observations.


\section{The CODEX models}\label{CODEXsection}
\begin{figure*}
 \sidecaption
   \includegraphics[width=12cm]{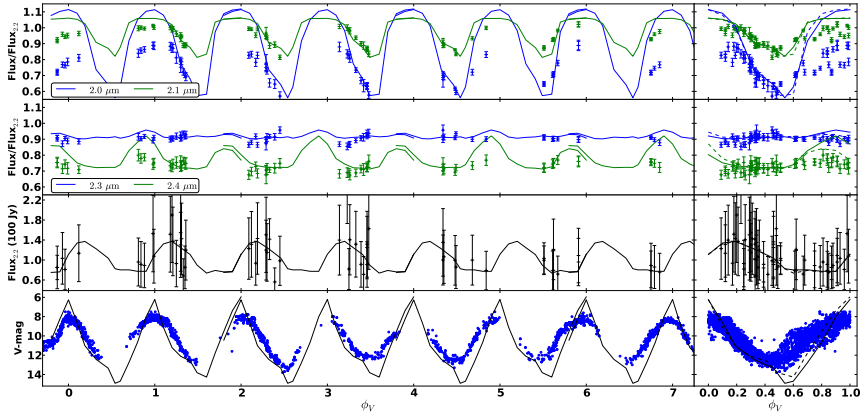} 
     \caption{Spectro-photometry. The two upper panels show the variation in the spectral shape with respect to the 2.2~$\mu$m channel, the third panel shows the absolute
              flux in the latter, and the lower panel the visual flux in magnitudes. In each panel two cycles of the R52 model series are consecutively 
              overplotted as the solid line. On the left, the cycles are shown as a function of visual phase while on the right they are phase-folded.}
     \label{spectra}
\end{figure*} 

The CODEX atmosphere models \citep[][]{Ireland2008MNRAS} are the most advanced publicly available dynamical models for M-type Mira variables. The pulsations 
in these models are self-excited and limited in amplitude by the non-linear effects of shocks in the outer atmosphere and by turbulent viscosity. For each model 
series, a few typical cycles were selected for detailed radiative transfer calculations with an opacity-sampling method in LTE with a phase-resolution of $\sim0.1$. 
This results in the centre-to-limb variation (CLV) and spectrum, with $\Delta \lambda = 0.0002 \, \mu $m between $\lambda = 0.45-2.50 \, \mu $m. 
The current model series are made to match o Cet, R Leo, and R Cas, although even these well-observed Miras have large uncertainties on their basic
parameters (e.g. the 2 very different models for R Cas). They were presented to the community and compared to photometric 
and interferometric observations in ISW11. The model parameters are mass M, luminosity L, metallicity Z, mixing length parameter $\alpha_m$, and 
turbulent viscosity parameter $\alpha_{\nu}$. For a detailed justification of the current set of parameter values, we refer to ISW11. 

With a slightly meandering ($\pm$ 5\,d) pulsation period of $\sim$317~d, TU And is compared to the R52 ($M=1.1\,M_{\odot}$, $L=5200\,L_{\odot}$, $Z=0.02$, $\alpha_m=3.5$, $\alpha_{\nu}=0.25$) 
and o54 ($M=1.1\,M_{\odot}$, $L=5400\,L_{\odot}$, $Z=0.02$, $\alpha_m=3.5$, $\alpha_{\nu}=0.25$) model series, which have derived periods of 307 and 330\,d, respectively. 

When comparing a model series to observations, the only free parameter is the distance, or equivalently, the phase-averaged Rosseland angular diameter $\theta_R$. Considering
the angular resolution of PTI, we converted the model intensities to synthetic observables for $\theta_R \in [1.5,4.5]$~mas, in steps of 0.05~mas.
Converting the model intensities to the spectro-photometric observables was done with Eq.~\ref{equation1}. The squared visibilities are computed following
\citet{Wittkowski2004A&A}:
\begin{equation}\label{equation2}
 V^2(B/\lambda_{\mathrm{eff}}) = \frac{\int_{\lambda_l}^{\lambda_u} \|(\nu (B/\lambda) \cdot F) \otimes G\|^2 \lambda d \lambda}{\int_{\lambda_l}^{\lambda_u} \|(\nu (0) \cdot F) \otimes G\|^2 \lambda d \lambda}
\end{equation}   
with $F$, $G$, $\lambda_l$, and $\lambda_u$ as in Eq.~\ref{equation1}, and
\begin{equation}\label{equation3}
 \nu(B/\lambda) = \int I_{\lambda}^{\mu} J_0(\pi \theta_R q B/\lambda) q d q
\end{equation}   
the monochromatic Hankel transform of the intensity profile $I_{\lambda}^{\mu}$ at the baseline $B$ and wavelength $\lambda$.


\section{Results}
Figure~\ref{spectra} shows the resulting photometric time series of TU And as a function of phase, with two consecutive cycles of the R52 model series repeatedly 
overplotted. The two other available cycles of this series give qualitatively identical results, so are not shown. The Rosseland angular diameter is chosen to match the absolute 
flux at 2.2~$\mu $m, with a resulting value of $\theta_R = 2.20 \pm 0.15$~mas or $\mathrm{d}=1.13 \pm 0.07$~kpc. Although the error on the absolute calibration 
in this channel is rather large, the contamination from upper atmospheric molecular layers at these wavelengths is smallest. This channel thus provides 
the best view at the continuum photosphere in the spectral region where the bulk of the luminosity is emitted. Moreover, fixing the angular diameter 
like this gives a consistent picture in both the spectro-photometry and the interferometry.  
 
The R52 model series shows reasonable agreement with the observed visual light curve in the lower panel, considering that (1) the amplitude was not tuned to match this light curve
and (2) non-LTE effects play an important role at these wavelengths \citep[][]{Ireland2008MNRAS}, causing an increase in visual brightness of 0.5 to 1.0 magnitude. 
In case the higher value of this correction is valid, the perceived difference between model and observations could point to the presence of an excess in visual light at 
maximum phases together with a too large model amplitude.

The two upper panels show the variation in the spectral shape with time. A clear discrepancy occurs around times of visual maximum, then the model predicts too much flux. Moreover,
the amplitude of this excess shows an interesting wavelength-dependence: the strongest effect is observed at 2.0~$\mu $m, a smaller discrepancy at 2.1 and 2.4~$\mu $m and 
a good agreement at 2.3~$\mu$m. 

The visibility time series is presented in Fig.~\ref{v2mozaiek}, overplotted with the same cycles of the R52 model series. Each panel displays the visibilities
at a given phase, using different symbols for observations coming from different cycles (at least two per panel). The model reproduces the observations well at phases 0.3-0.6, 
but not around visual maximum. 
Comparison with other cycles does not give better results. In fact, this reveals that the model has larger
cycle-to-cycle variations at pre-maximum phases than are seen in the observations. A closer look at the wavelength dependence shows that the discrepancy is least pronounced
at 2.3 and 2.4~$\mu$m. Between 2.0 and 2.2~$\mu$m, the model shows very little wavelength-dependence, unlike the observations which show a clear decrease in visibility towards
2.0~$\mu$m.

This observation is supported by the more quantitative assessment presented in Figs.~\ref{chi2map} and~\ref{chi2map2} (available electronically only). Epochs with 
observations on at least two different baselines were selected, to resolve the degeneracy between CLV basic size and shape. Each of these data sets with good uv-coverage 
was compared to each model, independently of model phase and for a range of $\theta_R$. The figures show the resulting $\chi^2$-maps. In all panels
a curved valley shows up: every data set can be fitted with a model of any phase, albeit with a significantly larger diameter for visual maximum models. However, within 
these valleys a better $\chi^2$ is typically found with a model at the correct (photometric) diameter but with a phase nearer to visual minimum than the actual value
of the observations. In conclusion, we find that the models around visual maximum should resemble those of visual minimum more in CLV shape and wavelength-dependence.

Finally, we also compared the observations to the similar o54 model series, except for its luminosity. The results for o54 are alike and 
thus not shown here. The visibilities seem to match slightly better, but the light curve discrepancies become larger. 


\section{Discussion}

\begin{figure}
  \resizebox{\hsize}{!}{\includegraphics{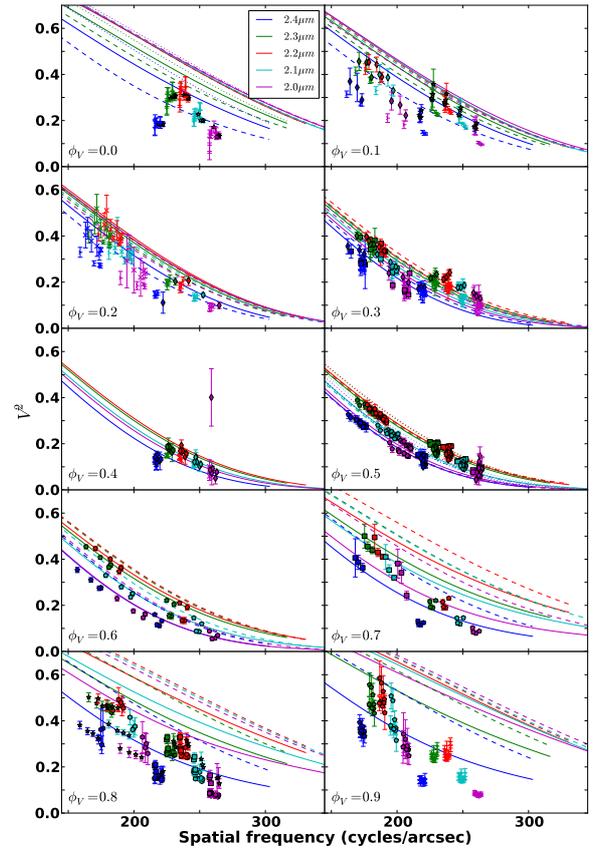}} 
     \caption{Visibility time series. Each panel shows all squared visibilities as a function of spatial frequency, obtained at the visual phase given in the 
              lower left corner of the panel. Different symbols are used for observations from different cycles. The same two model cycles as
              in Fig.~\ref{spectra} are overplotted as the full and dashed lines in each panel.}
     \label{v2mozaiek}
\end{figure}

\begin{figure}
   \resizebox{\hsize}{!}{\includegraphics{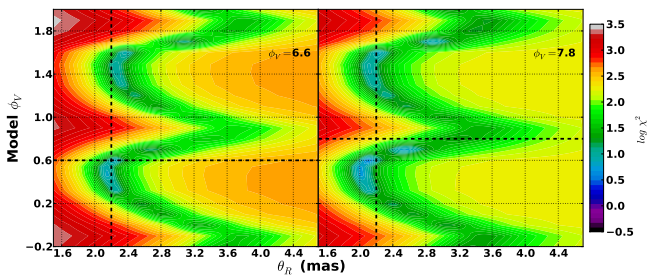}}   
     \caption{The $\chi^2$ maps of two data sets with a good uv-coverage (see text). The log $\chi^2$ is shown as a function of Rosseland angular diameter and
              phase, of the same two model cycles as in Figs.~\ref{spectra} and~\ref{v2mozaiek}. The phase of the data set is given in the upper right corner of 
              the panel and is plotted as the horizontal dotted line. The vertical dotted line marks the photometric diameter. The intersection of the dotted lines shows
              where the minimum $\chi^2$ should be if the model is correct.}
     \label{chi2map}
\end{figure}
 
\onlfig{4}{
\begin{figure}
   \resizebox{\hsize}{!}{\includegraphics{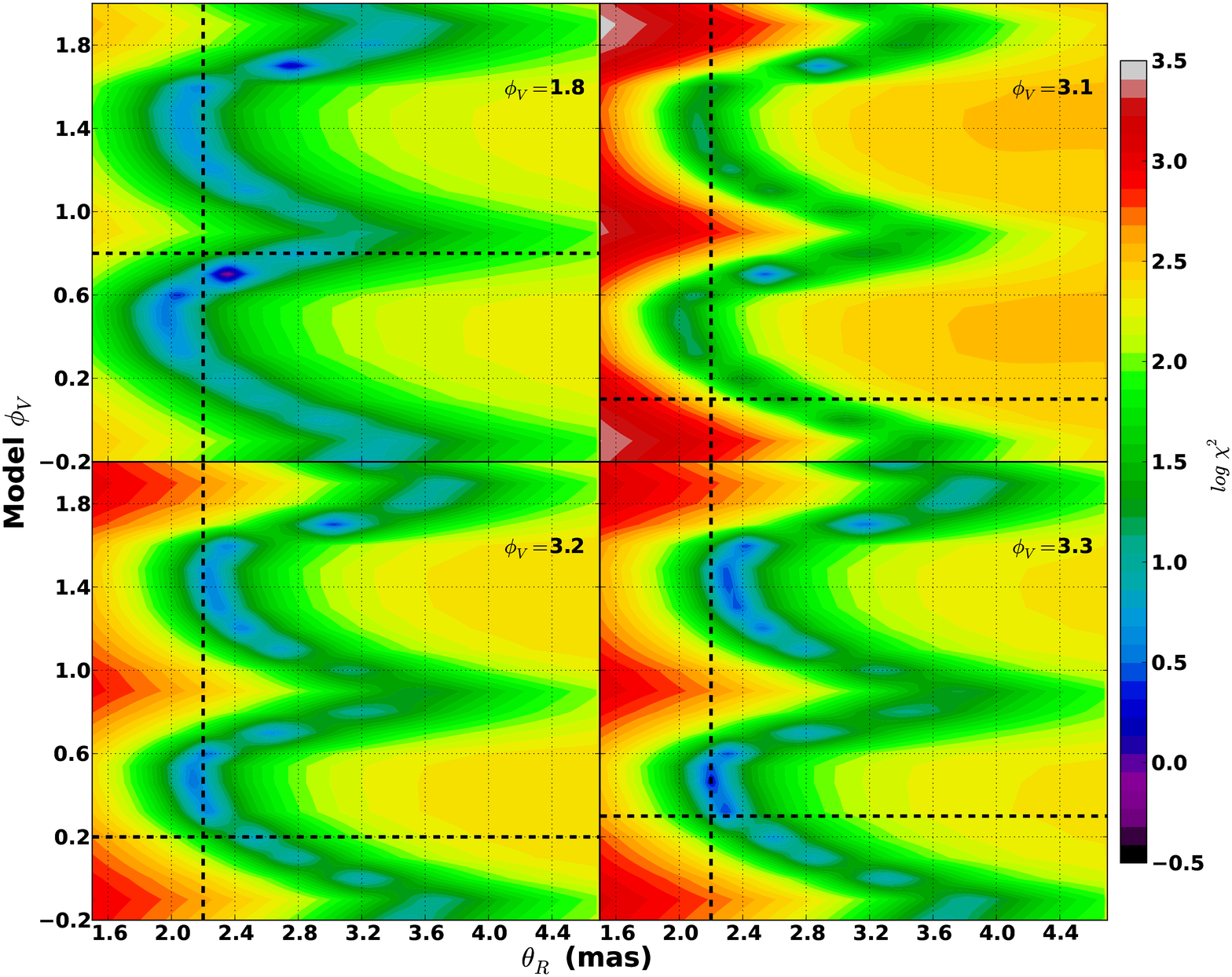}}   
     \caption{Similar to Fig.~\ref{chi2map}, but for the four other data sets with good uv-coverage. The log $\chi^2$ is shown as a function of Rosseland angular diameter and
              phase, of the same two model cycles as in Figs.~\ref{spectra} and~\ref{v2mozaiek}. The phase of the data set is given in the upper right corner of 
              the panel and is plotted as the horizontal dotted line. The vertical dotted line marks the photometric diameter. The intersection of the dotted lines shows
              where the minimum $\chi^2$ should be if the model is correct.}
     \label{chi2map2}
\end{figure} 
}

\citet{Ireland2008MNRAS} demonstrate that the CODEX models have reached a level of maturity, concerning computational issues and input physics, 
that allows quantitative comparisons with observations. Pulsations are self-excited and molecular layers are produced naturally and in a physically 
self-consistent way. 

Nevertheless, our observations of TU And show a clear discrepancy with the models around phases of visual maximum, both in the spectrum and in the visibilities. 
The most natural interpretation of the wavelength-dependence of this discrepancy seems to be in terms of the shape of the H$_2$O opacity curve, which shows an increase 
from 2.2~$\mu $m towards 2.0 and 2.4~$\mu $m. The formation of H$_2$O layers depends sensitively on the density, temperature, and pressure stratification in the upper atmosphere. 
In the models, the conditions in the upper atmosphere are set by the dynamics. Shock fronts that emerge from the continuum photosphere 
between phases -0.3 and -0.1 and then propagate outwards lead to the formation of an H$_2$O layer around phase 0.2-0.3, 
which then strongly decreases in strength at phase 0.7-0.8. At $\phi_V = 0.5$ both visibilities and spectra show good agreement. 
However, as demonstrated by Figs.~\ref{spectra},~\ref{v2mozaiek}, and~\ref{chi2map}, the near-maximum observations are represented better by model phases nearer to 
minimum, and thus with the presence of an H$_2$O layer in the atmosphere. This interpretation is supported by the better agreement at 2.3 and 2.4~$\mu $m. 
There CO is an important source of opacity, in the photosphere and in the extended atmosphere, even at visual maximum. 
As a much more stable molecule, CO depends less on the exact conditions and partly hides the lack of an H$_2$O molecular layer in the spectrum and visibilities.

In retrospect, it is interesting to note that the image reconstruction of T Lep \citep[P$\sim$380d, ][]{LeBouquin2009A&A}, made at a phase of 0.8-0.9, shows the clear signature 
of an extended molecular layer with a brightness ratio of $\sim$10\%, unlike its closest model series o54, which supports our findings. \citet{Woodruff2009ApJ} found fair
agreement between the wavelength dependence of their measured 1.1-3.8~$\mu$m UDs and the predictions of the CODEX models, for several Miras at phases 0.5 to 0.8, but needed
slightly larger model continuum diameters than predicted. On the other hand, \citet{Wittkowski2011A&A} have
recently compared AMBER medium spectral resolution observations of four Miras to the CODEX models and found good agreement, although they compared the models to stars of a different
period, hence luminosity. 

In summary, at visual maximum the current models predict a continuum photosphere that is too compact and hot as already suggested by ISW11, 
in combination with too small an H$_2$O optical depth in the outer atmosphere, as found here. 

The basic question now is, can this be salvaged within the current modelling framework? Is it just a question of tuning the parameters to more realistic values? 
Following the discussion of ISW11, it seems difficult to find any combination of parameter values that fits all observables. Accepting the LMC 
period-luminosity relation and assuming that the effect of metallicity is insufficient (see ISW11), the only free parameters are $M$, $\alpha_m$, and $\alpha_{\nu}$. Since the 
last mainly influences the pulsation amplitude, but not the period or the `parent star' radius, the above authors conclude that the problem at 
hand is one of calibrating the mixing length parameter as a function of mass. Decreasing the rather high $\alpha_m = 3.5$ to a more `realistic' value 
requires an increase in mass or a decrease in metallicity (albeit with a smaller dependency) to preserve the period, which would in case of o Cet 
be inconsistent with its kinematically determined age. TU And though, was classified as a member of a population with initial masses in the 
$1.15-1.4\,M_\odot$ range and $Z\in[0.004,0.02]$ \citep[][]{Mennessier2001A&A}, which makes it a better candidate to test the above hypothesis. 

We can at this point not gauge the impact of using an $\alpha_{\nu}$ which is not tuned to the amplitude of TU\,And. 
As a result, we feel that the model grid should span some scope in this parameter to allow more comparisons with `small-amplitude' pulsators.

Could either change in the model parameters also induce a stabilizing effect on the H$_2$O layer, as required by our observations, or is there 
a need to add more physics to the models?


\section{Conclusions}
The combination of a spectro-photometric and interferometric time series is a powerful tool for testing the validity of pulsating AGB model atmospheres, even at low spectral
resolution. PTI observed a large sample of M-type Miras and semi-regulars, several with a similar sampling to the one presented here, covering periods between 150 and 470d. 
To fully exploit this rich database, better coverage in the model parameter space is urgently needed, even though it requires a large computational effort. Only
in this way can the model parameters be calibrated and (simply) linked to Mira observable properties.
From the observational point of view, interferometric studies with better uv-coverage and spectral resolution should focus on phases 0.0 to 0.15, since none of the 
current model series predicts any extended molecular atmosphere in that phase range.
 
\begin{acknowledgements}
The Palomar Testbed Interferometer was operated by the NASA Exoplanet Science Institute 
and the PTI collaboration. It was developed by the Jet Propoulsion Laboratory, California Institute of Technology with funding 
provided from the National Aeronautics and Space Administration.
We acknowledge with thanks the variable star observations from the AAVSO International Database contributed by observers worldwide and used in this research.
\end{acknowledgements}

\bibliographystyle{aa}
\bibliography{aa_TUAnd}

\end{document}